\documentclass[preprint, prb, superscriptaddress]{revtex4}
\usepackage[dvips]{graphicx}
\begin{document}

\title
{
Estimation of spin-diffusion length 
from the magnitude of spin current absorption
}

\author
{ 
T. Kimura 
}
\affiliation{
Institute for Solid State Physics, University of Tokyo 
5-1-5 Kashiwanoha, Kashiwa, Chiba 277-8581, Japan
}
\affiliation{
RIKEN FRS, 2-1 Hirosawa, Wako, Saitama 351-0198, Japan 
}
\affiliation{
CREST, JST, Honcho 4-1-8, Kawaguchi, Saitama, 332-0012, Japan 
}
\email{kimura@issp.u-tokyo.ac.jp}

\author
{ 
J. Hamrle
}
\affiliation{
RIKEN FRS, 2-1 Hirosawa, Wako, Saitama 351-0198, Japan 
}
\affiliation{
CREST, JST, Honcho 4-1-8, Kawaguchi, Saitama, 332-0012, Japan 
}

\author{
Y. Otani
}
\affiliation{
Institute for Solid State Physics, University of Tokyo 
5-1-5 Kashiwanoha, Kashiwa, Chiba 277-8581, Japan
}
\affiliation{
RIKEN FRS, 2-1 Hirosawa, Wako, Saitama 351-0198, Japan 
}
\affiliation{
CREST, JST, Honcho 4-1-8, Kawaguchi, Saitama, 332-0012, Japan 
}


\date{\today}
\begin{abstract}
We demonstrate the method to calculate the spatial distributions 
of the spin current and accumulation 
in the multi-terminal ferromagnetic/nonmagnetic 
hybrid structure using an approximate electro-transmission line.  
The analyses based on the obtained equation yield the results 
in good agreement with the experimental ones.  
This implies that the method allows us to determine 
the spin diffusion length of additionally connected electrically 
floating wire from the reduction of the spin signal.

\end{abstract}

\maketitle

\section{Introduction}
Spin-dependent electron transport phenomena in nanostructured 
ferromagnet(F)/normal metal(N) hybrid systems show intriguing characteristics 
in association with the spin transfer and accumulation.  
The devices based on spin transport phenomena 
have great advantages over the conventional electronic devices 
because of additional spin functionalities.\cite{Bauer}  
Therefore, understanding responsible physics on the spin diffusion process 
becomes essential to realize spin-electronic devices.  
Especially, electrical spin injection from F 
into N, semiconductor or superconductor 
is an important technique for developing such devices.\cite{Schmidt}  
When a spin-polarized current flows across the junction between F and N, 
the spin-splitting in the chemical potential is induced in the N layer 
because of the sudden change in 
spin dependent electrical conductivity.\cite{vanSon}  
This leads to the spin accumulation in the vicinity of the F/N interface.    
The injected spin can be flipped due to 
spin-orbit interaction, magnon, phonon scattering and etc.  
The length scale over which the traveling electron spin memorizes the initial direction 
is known as the spin diffusion length, 
an important measure to realize the efficient spin injection.

First electrical spin injection into N 
was demonstrated by Johnson and Silsbee using a single-crystalline 
Al bulk bar.\cite{Johnson}  
They found that the Al bar has a long spin diffusion length 
of a few hundred microns.  
However, the obtained output spin signal was quite small in the 
range of nano ohms due to large sample dimensions.  
Current perpendicular to plane (CPP) giant magnetoresistance measurements 
in magnetic multilayers of alternating F and N layers 
enable us to perform more detailed study on the spin-diffusion process 
in the N layer because the relevant scale of the spin-diffusion length 
can be controlled by the spacer thickness.\cite{Bass1}  
This technique is suitable for the N layer with magnetic or 
even non-magnetic impurities 
which reduce the spin diffusion length dramatically.\cite{Bass2}  
However, for Ns in which long spin diffusion 
lengths are expected, the precise estimation of the spin diffusion length is 
unattainable because of difficulty in preparing CPP device with 
the thickness of the N spacer as thick as the spin diffusion length.  
On the contrary, planar mesoscopic spin-valve device 
is suitable to study the spin dependent electron transport 
of the material which has long spin diffusion length.  
Recently, Jedema {\it et al.} succeeded in detecting the clear spin signal 
in the lateral structure 
even at room temperature by using nonlocal spin-valve (NLSV) 
measurements similar to the Johonson's potentio-metric method.\cite{Jedema}  
In order to detect clear spin signals in the NLSV measurements, 
the spacing between the F injector and detector 
should be shorter than the spin-diffusion length of the material.  
This means that, for the Ns 
with spin diffusion length of sub hundred nanometers, 
the NLSV technique cannot be applied 
because of the technical limit of device fabrication.  
We have experimentally demonstrated that in such lateral structures 
an additional floating probe which does not carry the charge current, 
affects the spin transport when the probe is located 
within the spin diffusion length from the spin injector.\cite{Kimura} 
In this article, we further extend the experiment for quantitative estimation of 
the spin-diffusion length using the spin absorption effect. ({\it i.e.} spin sink)

\section{Basic equations}
The continuous charge current $I_C$ in the F subsection 
provides the spin current $I_S = \alpha_F I_C$, 
with $\alpha_F$ of the spin polarization in the F subsection.  
When the spin current flows across the F/N junction, 
the spin splitting in the chemical potential is induced in both 
F and N subsections, as shown in Fig.\ 1(a).   
Here, we devide the spin splitting chemical potential into two components.  
One is the voltage drop due to the charge current 
with the ohmic resistance in each layer shown in Fig. 1(b).  
In this case, the distribution of the chemical potential is obtained by 
considering the conventional electrical circuit consisting of a series jucntion 
with two different resistances.  
A constant spin current $I_S$ is induced by the continuous charge current 
in the F subsection.  However, no spin current exists in the N subsection.  
The other one is the spin dependent the chemical potential 
without the contribution of ohmic-resistance voltage drop as shown in Fig. 1(c), 
where spin currents are generated both in the N and F layers.  
Since the sorce of the induced spin currents $I_S$ is originated from the charge current of the F layer, 
the relation $I_S = I_{{\rm SF}} + I_{{\rm SN}}$ is satisfied at the interface.  
Here $I_{{\rm SF}}$ and $I_{{\rm SN}}$ are respectively the spin currents in the F and N subsections 
induced by the spin splitting of the chemical potential.  
Thus, the F/N junction with the charge current $I_C$ acts as 
a spin-current source with the magnitude of $\alpha_F I_C$.

We consider the spin-splitting component of the chemical potential 
in which there is no charge current.  
The essential features for the spin-current and spin-accumulation distributions 
in the diffusive regime are described 
by the spin-dependent Boltzman equation.\cite{Fert1}  
In general, from the one-dimensional diffusion equation, 
the induced spin-splitting voltage 
$\Delta V_{\rm S} = \Delta \mu/e = (\mu_{\uparrow} - \mu_{\downarrow})/e$, 
where $\mu_{\uparrow}$ and $\mu_{\downarrow}$ are the chemical potential 
of the up- and down- spins, respectively, is given by 
\begin{equation}
\Delta V_{\rm S} = V_+  e^{-\frac{x}{\lambda}} + V_- e^{\frac{x}{\lambda}}.  
\end{equation}
Here $\lambda$ is the spin diffusion length.  
In the N layer, using the spin-dependent ohmic law $ I_{\uparrow, \downarrow}
= (-\sigma_{\rm N} S/2) \partial V_{\uparrow, \downarrow} /\partial x$, 
the spin current $I_{{\rm SN}} = I_{\uparrow} - I_{\downarrow}$ is calculated as 
\begin{equation}
I_{{\rm SN}}  =  \frac{\sigma_{\rm N} S_{\rm N}}{2 \lambda_{\rm N}} 
(V_+  e^{-\frac{x}{\lambda_{\rm N}}} - V_- e^{\frac{x}{\lambda_{\rm N}}}) 
 =  \frac{1}{R_{{\rm SN}}} (V_+  e^{-\frac{x}{\lambda_{\rm N}}} - 
V_- e^{\frac{x}{\lambda_{\rm N}}}), 
\end{equation}
where, $S_{\rm N}$, $\sigma_{\rm N}$ and $\lambda_{\rm N}$ are 
the cross sectional area, the conductivity 
and the spin diffusion length  of the N subsection, respectively.  
The spin resistance 
$R_{{\rm SN}}$ is a measure of the difficulty for spin mixing  
and is defined as $2 \lambda_{\rm N}/(\sigma_{\rm N} S_{\rm N})$.  
The situation described by Eqs.\ 1 and 2 is 
equivalent to the electrical transmission line of 
the characteristic impedance $R_{{\rm SN}}$ with 
the attenuation constant $1/\lambda_{\rm N}$.\cite{Ramo}

We can also obtain the similar relation for the F layer.  
In the F subsection, the spin current $I_{{\rm SF}}$ is given by 
\begin{eqnarray}
I_{{\rm SF}} & = & \frac{\sigma_{\rm F} S_{\rm F}}{2} \left( (1+\alpha_{\rm F}) 
\frac{\partial V_{\uparrow}}{\partial x} 
-  (1-\alpha_{\rm F}) \frac{\partial V_{\downarrow}}{\partial x}  \right)\\
   & = & \frac{\sigma_{\rm F} S_{\rm F}}{2} \left( \frac{\partial \Delta V_{\rm S}}{\partial x} + 
   \alpha_{\rm F} \left 
   (\frac{\partial V_{\uparrow}}{\partial x} + \frac{\partial V_{\downarrow}}{\partial x} 
   \right)
   \right)
\end{eqnarray}
where, $S_{\rm F}$ and $\sigma_{\rm F}$  are 
the cross sectional area and the conductivity in the F subsection, respectively.  
Since there is no charge current, the relation 
$ \alpha_{\rm F} \partial \Delta V_{S}/\partial x 
=  -(\partial V_{\uparrow}/\partial x 
+ \partial V_{\downarrow}/\partial x)$ is satisfied.  
Therefore, the spin current $I_{{\rm SF}}$ in the F layer is given by 
\begin{equation}
I_{{\rm SF}}  =  \frac{(1-\alpha_{\rm F}^2)\sigma_{\rm F} S_{\rm F} }{2\lambda_{\rm F}} 
(V_+  e^{-\frac{x}{\lambda_{\rm F}}} - 
V_- e^{\frac{x}{\lambda_{\rm F}}}) 
 =  \frac{1}{R_{{\rm SF}}} (V_+  e^{-\frac{x}{\lambda_{\rm F}}} - 
V_- e^{\frac{x}{\lambda_{\rm F}}}), 
\end{equation}
where, $\lambda_{\rm F}$ is the spin diffusion length in the F subsection.  
Thus, in the F subsection, $R_{{\rm SF}}$ is given 
by $(1/(1-\alpha_{\rm F}^2))(2\lambda/(\sigma_{\rm F} S_{\rm F}))$ 
instead of $2\lambda_{\rm N}/(\sigma_{\rm N} S_{\rm F})$.  
Eqs. 2 and 5 mean that the characteristic spin resistance over the spin-diffusion length 
is comparable to the characteristic impedance for the spin current and spin-splitting voltage.

Using these relations, we can simplify 
the calculation of the spatial distributions of 
spin current and accumulation in complex multi-terminal F/N hybrid structures
consisting of different characteristic spin resistances.  
We will show that the characteristic spin resistance $R_S$ 
is an important measure to design the spin-dependent-transport property of the system.

First, we consider a cascaded transmission line 
of two different characteristic spin resistances as shown in Fig.\ 2(a).  
Here, the first and second subsections have 
the characteristic resistance $R_{\rm S1}$ with 
the spin-diffusion length $\lambda_{1}$ and 
the characteristic resistance $R_{\rm S2}$ with 
the spin diffusion length $\lambda_{2}$, respectively.  
The length of the first subsection is $d$ 
and the second subsection extends to infinity.  
The basic equations $\Delta V_{\rm Si} (x) 
= V_{i+} e^{-x/\lambda_i} + V_{i-} e^{x/\lambda_i}$ 
and $I_{Si} (x) = (V_{i+} e^{-x/\lambda_i} - V_{i-} e^{x/\lambda_i})/R_{Si}$ 
are valid for each subsection.  
In the second line, the coefficient $V_{2-}$ of $e^{x/\lambda_2}$ 
is zero because of its infinite length.  
Therefore, we obtain $\Delta V_{\rm S2} (x) = V_{2+} e^{-x/\lambda_2}$ and 
$I_{\rm S2} (x) = \Delta V_{\rm S2}(x)/R_{\rm S2}$.  
This relation and the continuities of $I_S$ and 
$\Delta V_S$ at the junction yield 
$V_{1+}=e^{d/\lambda} V_{2+} (R_{\rm S1} +R_{\rm S2})/2R_{\rm S2}$ 
and $V_{1-}=e^{-d/\lambda}V_{2+} (R_{\rm S2}-R_{\rm S1})/2R_{\rm S2}$.  
We should note that the situations where $V_{1-} > 0$ and $V_{1-} < 0$ 
respectively correspond to 
the spin-current absorption into the second line  
-and the reflection from it 
due to the resistance mismatch at the boundary.  
We then define the transmission coefficient $T$ as the ratio of 
the spin splitting voltage $V_{\rm S1}$ at the junction 
to the voltage $V_{\rm S0}$ at the input terminal.   
The transmission coefficient $T \equiv \Delta V_{\rm S1}/\Delta V_{\rm S0}$ 
is given by 
\begin{equation}
T   =  
\frac{R_{\rm S2}}{R_{\rm S1} \sinh (d/\lambda_1) + R_{\rm S2} \cosh (d/\lambda_1) } 
 =  \frac{Q}{\sinh (d/\lambda_1) + Q \cosh (d/\lambda_1)},  
\label{Trans}
\end{equation} 
where, $Q$ is the ratio $R_{\rm S2}/R_{\rm S1}$.  
Defining the input spin resistance as $\Delta V_{\rm S0} /I_{\rm S0}$ 
where $I_{\rm \rm S0}$ is the spin current at the input terminal, 
We obtain the input spin resistance in the series connection 
$R_{\rm S-{\rm Series}}$ as 
\begin{equation}
R_{\rm S-{\rm Series}}  =  \frac{R_{\rm S1}\sinh (d/\lambda_1)+R_{\rm S2}\cosh (d/\lambda_1)}
{R_{\rm S1}\cosh (d/\lambda_1) +R_{\rm S2}\sinh (d/\lambda_1)} R_{\rm S1} 
 =  \frac{\sinh (d/\lambda_1) + Q \cosh (d/\lambda_1)}
{\cosh (d/\lambda_1) + Q \sinh (d/\lambda_1)} R_{\rm S1}.  
\label{Series}
\end{equation}
The input spin resistance $R_{\rm S-{\rm Parallel}}$ in a parallel connection 
shown in Fig.\ 2(b) is also calculated as 
\begin{equation}
R_{\rm S-{\rm Parallel}}  =  \frac{R_{\rm S1} R_{\rm S2}}{R_{\rm S1}+R_{\rm S2}} \\ 
       =  \frac{R_{\rm S2}}{1+Q}.
\label{Parallel}
\end{equation}
Above basic equations can be applied for 
calculating the characteristics in multi-terminal spin-transport systems.

In the above calculation, we assume a transparent interface between 
the different subsections with zero resistance.  
This assumption is justified only in limited materials.    
However, in this article, we treat Cu/Py and Cu/Au
 interfaces.  
These interfaces are known to have small resistance 
compared with other interfaces such as Cu/Pd and Cu/Pt.\cite{Bass1, Bass2, Bass4}  
Therefore, we neglected the interface resistance.

\section{Results and Discussions}
\subsection{Single junction}
We now consider a single F/N cross junction with 
the charge current $I_C$ flowing across the F/N interface 
as shown in Fig.\ 2(c).  
As mentioned above, the charge current across the junction 
plays a role of the spin current source with the magnitude of $\alpha_F I_C$.  
The induced spin current diffuses into each subsection 
according to the magnitude of characteristic spin resistance. 
This can be described as the parallel spin resistance circuit 
consisting of two F and two N subsections shown in Fig.\ 2(c).  
We can now calculate the induced spin-splitting voltage $V_{\rm S0}$ at the interface as 
\begin{equation}
\Delta V_{\rm S0} = \alpha_F I_C \frac{R_{\rm SF} R_{\rm SN}}{2(R_{\rm SF} + R_{\rm SN})}.
\end{equation}
The spin splitting decays exponentially 
over the spin diffusion length in each subsection.

\subsection{Double junction}
Unfortunately, we cannot experimentally determine the spin splitting of the chemical potential 
in the above single junction 
although the spin splitting can be induced in the N subsection.  
When an additional F probe (${\rm F_2}$) is connected to the N 
subsection within the spin diffusion length 
from the spin-polarized current injector
as shown in Fig.\ 2(d), 
the voltage difference between the ${\rm F_2}$ 
and N subsections can be detected as the boundary resistance 
caused by the spin splitting.  
Remarked here is that the spin-current 
and chemical-potential distributions are significantly affected 
by the additional wire.  
We discuss the double F/N junctions on the basis of 
the transmission model.  
In this case, the input spin resistance $R_{\rm S{\rm i}}$ for 
the nonlocal lead consisting of the N and F layers 
changes from $R_{\rm S{\rm N}}$, 
resulting in the different spin-splitting voltage from the single junction.  
The $R_{\rm S{\rm i}}$ can be calculated by considering 
a cascaded transmission line 
of $R_{\rm S{\rm N}}$ and $R_{\rm Sd}$ with the junction at the position $d$.  
Here, $R_{\rm Sd}$ is the spin resistance from the detector junction 
and is given by 
\begin{equation}
R_{\rm Sd} = \frac{R_{\rm SN} R_{\rm SF}/2 }{R_{\rm SF}/2 + R_{\rm SN}} = \frac{R_{\rm SF}}{2+Q}.  
\end{equation}
From Eq. \ref{Series}, $R_{\rm S{\rm i}}$ is deduced as 
\begin{equation}
R_{\rm S{\rm i}}  =  \frac{R_{\rm SN}\sinh (d/\lambda_N)+R_{\rm Sd}\cosh (d/\lambda_N)}
{R_{\rm SN}\cosh (d/\lambda_{\rm N}) +R_{\rm Sd}\sinh (d/\lambda_{\rm N})}
 R_{\rm SN} 
 =  R_{\rm SN} - \frac{2\,R_{\rm SN}}
   {1 + e^{\frac{2\,d}{\lambda_{\rm N}}}\,\left( 1 + 2 Q \right) }.  
\end{equation}
The total spin resistance of the device $R_S$ is given as 
the parallel sum of $R_{\rm N}$, two $R_{\rm F}$ and $R_{\rm Si}$ : 
\begin{equation}
R_{\rm S}  =  \left( \frac{1}{R_{\rm SN}} + 
    \frac{2}{R_{\rm SF}} + \frac{1}{R_{\rm Si}}
\right)^{-1}  
 =  \frac{Q R_{\rm SN}
     \left( e^{\frac{d}{\lambda_{\rm N}}}Q + 2\sinh(d/\lambda_{\rm N}) 
\right) }
{2 \left( e^{\frac{d}{\lambda_{\rm N}}} Q \left( 2 + Q \right)  + 
     2\sinh (d/\lambda_{\rm N}) \right) }.
\end{equation}

Also using Eq. \ref{Trans}, the transmission coefficient of the device $T_2$ 
corresponding to the ratio of the spin-splitting voltage 
between the injector and detector F/N interfaces can be calculated as 
\begin{equation}
T_2  =  \frac{R_{\rm Sd}}{R_{\rm SN} \sinh (d/\lambda_{\rm N}) + R_{\rm Sd} 
\cosh (d/\lambda_{\rm N})} 
   =  \frac{Q}{e^{\frac{d}{\lambda_{\rm N}}}\,Q + 2\,\sinh (d/\lambda_{\rm N})}
\end{equation}
The induced spin-splitting voltage at the injector junction is given by $I_S R_S$.  
Then, the induced spin-splitting voltage $\Delta V_{\rm S2}$ at the detector junction 
can be calculated by $\Delta V_{\rm S2} = T_2 I_S R_S$.  
Since the NLSV signal $\Delta V_{\rm \rm N2}$ is given by $\alpha_F \Delta 
V_{\rm S2}$\cite{Kimura2}, 
we obtain   
\begin{equation}
\Delta V_{\rm N2} = \frac{\alpha_F I_S Q^2 R_{\rm Sn}}
   {2 e^{\frac{d}{\lambda_N}}\, \left( 2 + Q \right)  + 
     4 \sinh (d/\lambda_N) }.  
\label{double}
\end{equation}

We experimentally demonstrate the precise estimation of the spin diffusion lengths 
of F and N strips and the spin polarization of F strips 
using Eq. \ref{double}.  
The lateral spin valves for this study 
consist of double Py/Cu junctions.  
Figure 3(a) shows a SEM image of the typical fabricated device.  
Here, the width of both Py and Cu strips is 100 nm and 
the thicknesses of Py and Cu strips are 30 nm and 80 nm, respectively.  
The resistivities of the Py and Cu strips are respectively 26.8 $\mu\Omega$cm 
and 2.08 $\mu\Omega$cm at room temperature.  
We vary the distance between the injector and detector junctions 
from 270 nm to 700 nm and measure the NLSV signal 
at room temperature as a function of the distance.  
In order to compare quantitatively 
Eq. \ref{double} with the experimental spin signal, 
the cross sections $S_{\rm \rm Cu}$ and $S_{\rm \rm Py}$ 
have to be known to obtain the spin resistance of each strip.  
We thus have to carefully estimate the cross sections.  
From previous reports, $\lambda_{\rm \rm Cu}$ is of the order 
of several hundred nanometer 
and $\lambda_{\rm \rm Py}$ is a few nanometers.\cite{Jedema, Bass3, Albert, Fert2}  
Since the spin-current flows along the Cu strip 
over a few hundred of nanometers
the area $S_{\rm \rm Cu}$ for estimating $R_{\rm \rm Cu}$ 
should be given by the cross section of the Cu strip (100 nm $\times$ 80nm).  
On the other hand, in the Py strip, 
the cross section is not appropriate 
for the area $S_{\rm \rm Py}$ for $R_{\rm S{\rm Py}}$ 
because the spin current diminishes in the vicinity of the junction.  
Therefore, for the Py strip $S_{\rm \rm Py}$ should be 
the junction-area 100 nm $\times$ 100 nm. \cite{Takahashi}

A NLSV signal at a distance of 270 nm as a function of the external magnetic field 
parallel to the Py strip is shown in the inset of Fig.\ 3(b).  
We observe the clear spin-accumulation signal at room temperature.  
Figure 3(b) shows the NLSV signal as a function of the distance $d$.  
The obtained signal decreases monotonically 
with increasing the distance $d$ due to the spin relaxation.  
This result is fitted to Eq. \ref{double}.  
As shown in the Fig.\ 3(b), 
the fitted curve is in good agreement with the experimental results.  
From the fitting parameter, 
we obtain the spin-diffusion length of the Cu strip 
and that of the Py strip as 500 nm and 3 nm, respectively.   
And also, the polarization $\alpha_{\rm \rm Py}$ of Py strip is 
determined as 0.25 at room temperature.  
Here, $\lambda_{\rm {\rm Cu}} = 500$ nm is quite long 
compared with other reported values.\cite{Jedema, Albert}  
We notice that the resistivity of our Cu 
is rather smaller than that of the reported one.  
This supports the long spin diffusion length of our Cu strip.  
The $\lambda_{\rm {\rm Py}} = 3$ nm and $\alpha_{\rm {\rm Py}}=0.25$ 
are reasonable although 
the values are slightly smaller than other group's values.  
This may be due to rather higher resistivity of our Py than others.  
Using these values, 
we can calculate the characteristic spin resistance of each strip 
as $R_{\rm S{\rm Cu}}=2.60$ $\Omega$ and $R_{\rm S{\rm Py}}=0.34$ $\Omega$.

\subsection{Triple junction}
We extend the above formalism to a triple junction in Fig.\ 2(e).  
In this case, the output spin-splitting voltage 
induced between the F2 and N strip by 
the current injection from F1 is influenced by the M strip 
which absorbs the induced spin current.  
The middle strip has the spin resistance of $R_{\rm S{\rm M}}$ 
with the spin diffusion length $\lambda_{\rm \rm M}$.  
The magnitude of the spin-current and spin-accumulation 
can be calculated by using the method explained above .   
From the calculation, 
the position of the middle wire was found not to affects the spin signal.  
Therefore, for simplicity we assume that 
the middle wire is located in the center between F1 and F2 strips 
which have the same spin resistances of $R_{\rm S{\rm F}}$.  
When the current is injected from F1, 
the NLSV voltage $\Delta V_{\rm \rm N3}$ at junction of F2/N is 
given by 
\begin{equation}
\Delta V_{\rm \rm N3} = 
 \frac{\alpha_F I_S Q^2 Q_{\rm M} R_{\rm SN}}
  {2\left( -2\left( 1 + Q \right) + 
      \left( 2 + Q\left( 2 + Q \right) \,
          \left( 1 + Q_{\rm M} \right)  \right) 
       \cosh (\frac{d}{\lambda_{\rm N}}) + 
      \left( 2Q_{\rm M} + Q \left( 2 + Q \right) \,
          \left( 1 + Q_{\rm M}  \right)  \right) \,
       \sinh (\frac{d}{\lambda_{\rm N}}) \right) } , 
\end{equation}
where, $Q$ and $Q_{\rm M}$ 
are respectively defined as $R_{\rm S{\rm F}}/R_{\rm S{\rm N}}$ and $R_{\rm S{\rm M}}/R_{\rm S{\rm N}}$.   
$d$ is the distance between the injector and detector.  
$Q_{\rm M} = \infty$ corresponds the case without M strip.

The lateral spin valves 
consisting of the triple junctions 
are then prepared 
with changing the material for the M strip.  
We know the spin resistances 
of Py and Cu strips from the previous double-junction experiment.  
Therefore, the Py and Cu wires were used as F and N strip, respectively.   
Cu, Py and Au are chosen as the material of the middle strip.   
The width and thickness of each strip 
are the same as those of the double junction.  
Figure 4 shows a fabricated lateral spin valve of the triple junctions 
with a Au M strip.  
The distance $d$ between injector and detector Py/Cu junctions is 650 nm.  
The M wire is located 
at an intermediate position between the injector and detector.  
The NLSV is measured with the probe configuration in Fig.\ 4(b).  
Using the spin resistances $R_{\rm S{\rm Cu}} = 2.60 \Omega$, 
$R_{\rm S{\rm Py}} = 0.34 \Omega $ and Eq.\ \ref{NLSV}, 
we obtain the following equation between 
the detected spin signal $R_{\rm \rm NLSV}$ and $R_{\rm SM}$   
\begin{equation}
R_{\rm \rm SM} [\Omega] = 
\frac{7.02 R_{\rm \rm NLSV}[{\rm m} \Omega]} {1.27 - 4.41 R_{\rm \rm NLSV}[{\rm m} \Omega]} .  
\label{NLSV}
\end{equation}

Figure 5(a) shows the NLSV signal of the device with the Cu M strip.  
We obtained the spin signal of 0.18 m$\Omega$ 
smaller than the value of 0.27 m$\Omega$ 
obtained without the middle wire.  
This is because a small amount of the spin current 
is absorbed into the middle Cu strip.  
Figure 5(b) shows the NLSV signal of the device with the Py M strip.  
Here the difference in the switching field between Figs.\ 5(a) and (b) 
due to the shape of both injector- and detector-F strips 
to the large pad.\cite{Kimura} 
The spin signal exhibits a drastic reduction to 0.05 m$\Omega$, 
much smaller than that without the M strip.  
This is also due to the spin-current absorption 
into the Py M strip.  
However, because of the small spin resistance of the Py strip, 
the spin current is preferably absorbed into the Py M strip, 
resulting in much smaller spin signal in the NLSV measurement.    
Figure 5(c) shows the NLSV signal of the device 
with the Au M strip.  
The obtained spin signal is 0.08 m$\Omega$ and 
shows the large reduction of the spin signal 
similar to that of the M Py strip.  
This implies that the Au wire has smaller spin resistance.

Figure 6 shows the spin resistance of the M strip $R_{\rm \rm SM}$
as a function of $R_{\rm \rm NLSV}$ based on Eq. \ref{NLSV}.  
From the equation, we obtain the spin resistances of 
the Cu, Au and Py stirp as 2.67, 0.62 and 0.33 $\Omega$.  
The experimental values for the Py and Cu M strips 
are quantitatively in good agreement with the ones obtained from 
the double junction experiment.  
This indicates that Eq.\ \ref{NLSV} is valid 
for estimating the spin resistance of the M strip.  
For the Au strip, since the resistivity of the Au strip is 5.24 $\mu \Omega$cm, 
we obtain the spin-diffusion length of the Au strip as 60 nm.  
This is in good agreement with the other reports \cite{Bass3} 
and proves that the Au strip has short spin-diffusion length 
due to the strong spin-orbit interaction.\cite{Pannetier}  
Thus, using this method, 
we can estimate the characteristic spin resistance 
and spin diffusion length of the material.

\section{Conclusion}
We have shown the calculation of 
the spin-current and spin-splitting-voltage distributions 
in the multi-terminal F/N hybrid structure 
by introducing the characteristic spin resistance 
and similarly treating the system as an electro-transmission line.   
The analyses based on this method give 
the values in good agreement with the experimental results.  
The spin diffusion length of our Cu strip was found to 
be 500 nm quite long at room temperature.  
The model is extended to the triple F/N junctions 
consisting of a conventional lateral spin valve 
with an additionally connected floating strip 
which is located in between the injector and detector.  
When the spin resistance of the additional strip 
is much smaller than that of the N strip, 
the spin currents tend to flow into the additional strip to be relaxed.  
We also emphasize that, using the absorption effect, 
the spin-diffusion length of the additional strip 
can be estimated quantitatively from the reduction of the spin signal 
in the NLSV measurement.  
The situation of the present experiment may be directly compared with  
the spin pumping effect observed in magnetic multilayers.\cite{pump}

\newpage

\begin{figure}

\caption{
(a)  Schematic illustrations of 
the spin splitting chemical potential 
induced by spin injection in F/N junciton.  
(b) A component of the voltage drop due to the ohmic resistance 
and (c) that of the spin splitting without charge current.  
}

\caption{
(a)  Schematic illustrations of 
junction connected in series (a) 
and in parallel (b).  
with different characteristic 
spin resistances 
The distribution of the spin current and spin splitting 
can be similarly treated 
as an electrical transmission line.  
Schematic illustrations of (c) a single cross F/N junction with 
the equivalent circuit for the transmission calculation, 
(d) that of a double F/N junction 
and (e) that of triple junction.  
}

\caption{
(a)  SEM images of the typical lateral spin device consisting of 
the double Py/Cu junctions.  
(b) NLSV signal measured at room temperature as a function of 
the distance $d$ between F injector and F detector.  
The inset of (b) is a obtained NLSV curve for the distance $d =$ 270 nm.  
}

\caption{
SEM images of the typical lateral spin device consisting of 
the double Py/Cu junctions and the middle Au/Cu junction.  
}

\caption{
(a) Non-local spin-valve curve with the Cu middle wire, 
(b) that with the Py middle wire 
and (c) with for the Au middle wire.  
}

\caption{
Calculated  spin resistance as a function of 
the obtained non-local spin signal $R_{\rm \rm NLSV}$.  
}

\end{figure}

\vspace*{1cm}
\newpage
\vspace*{4cm}
\begin{center}
\includegraphics[scale=0.5]{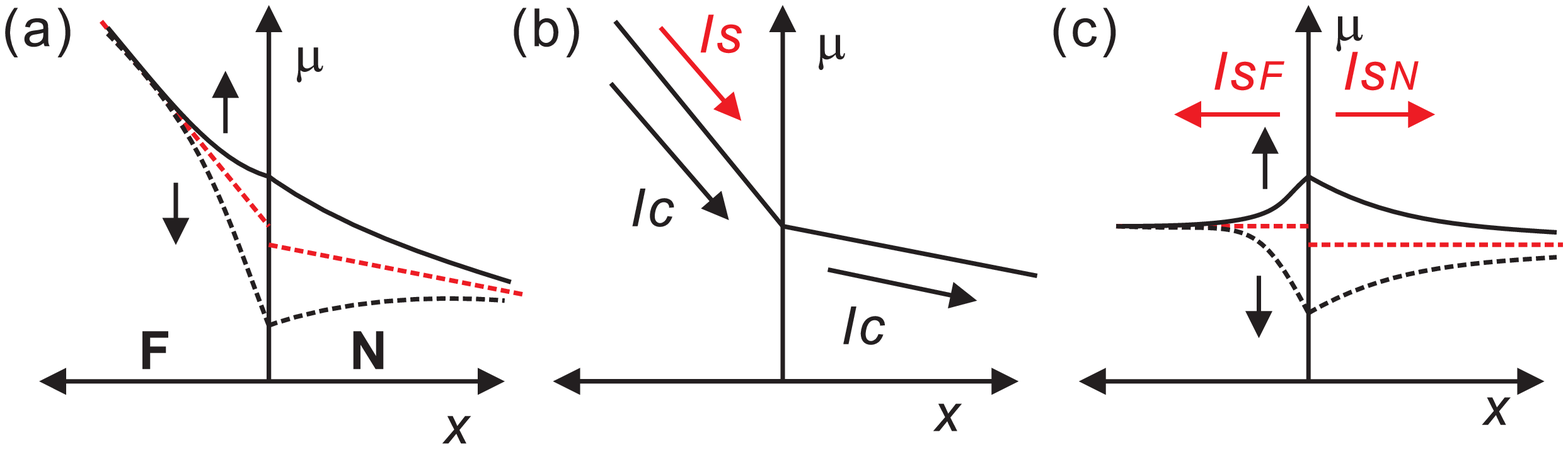}
\end{center}
\vspace*{1cm}
\begin{center}
Fig.\ 1 Kimura et al.
\end{center}

\newpage
\begin{center}
\vspace*{3cm}
\includegraphics[scale=0.8]{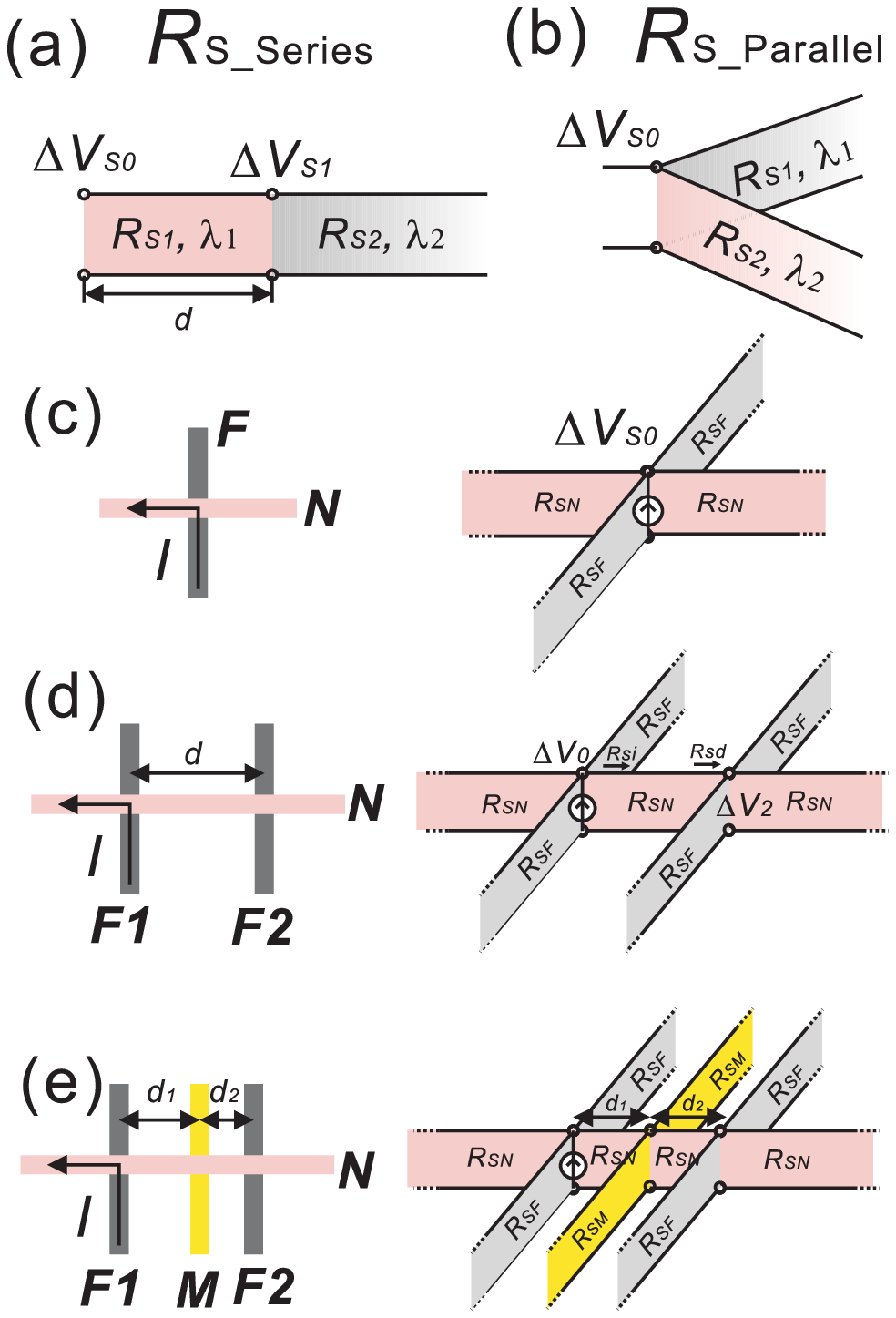}
\end{center}
\vspace*{1cm}
\begin{center}
Fig.\ 2 Kimura et al.
\end{center}

\newpage
\vspace*{4cm}
\begin{center}
\includegraphics[scale=0.8]{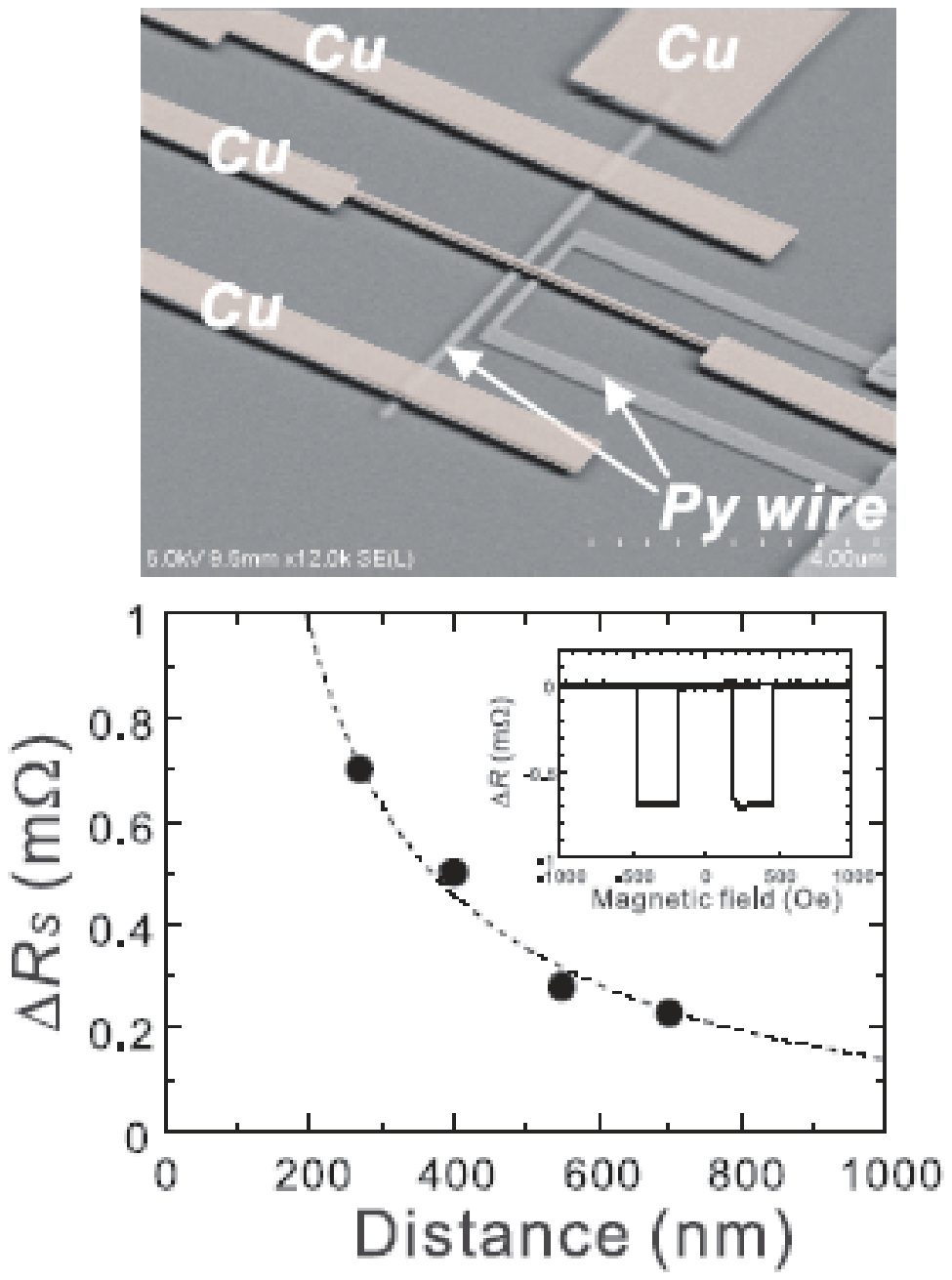}
\end{center}
\vspace*{1cm}
\begin{center}
Fig.\ 3 Kimura et al.
\end{center}

\newpage
\vspace*{4cm}
\begin{center}
\includegraphics[scale=0.6]{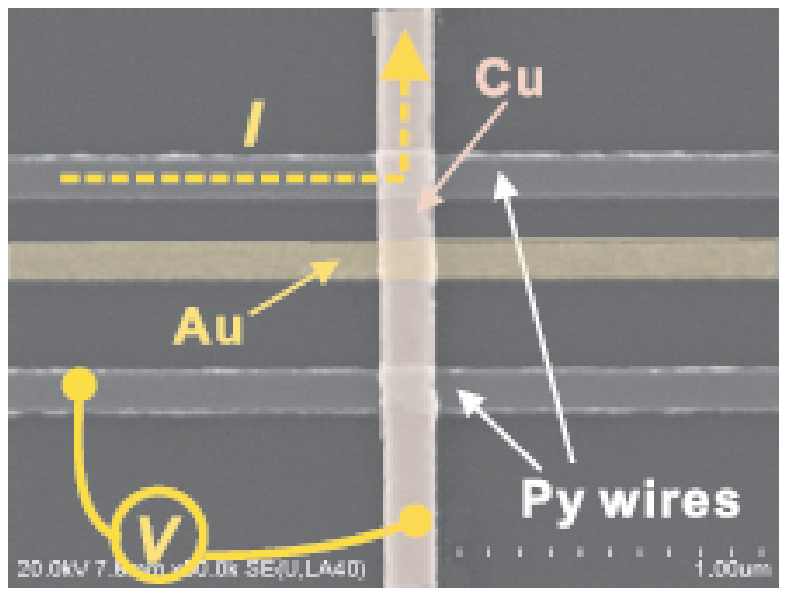}
\end{center}
\vspace*{1cm}
\begin{center}
Fig.\ 4 Kimura et al.
\end{center}

\newpage
\vspace*{2cm}
\begin{center}
\includegraphics[scale=0.7]{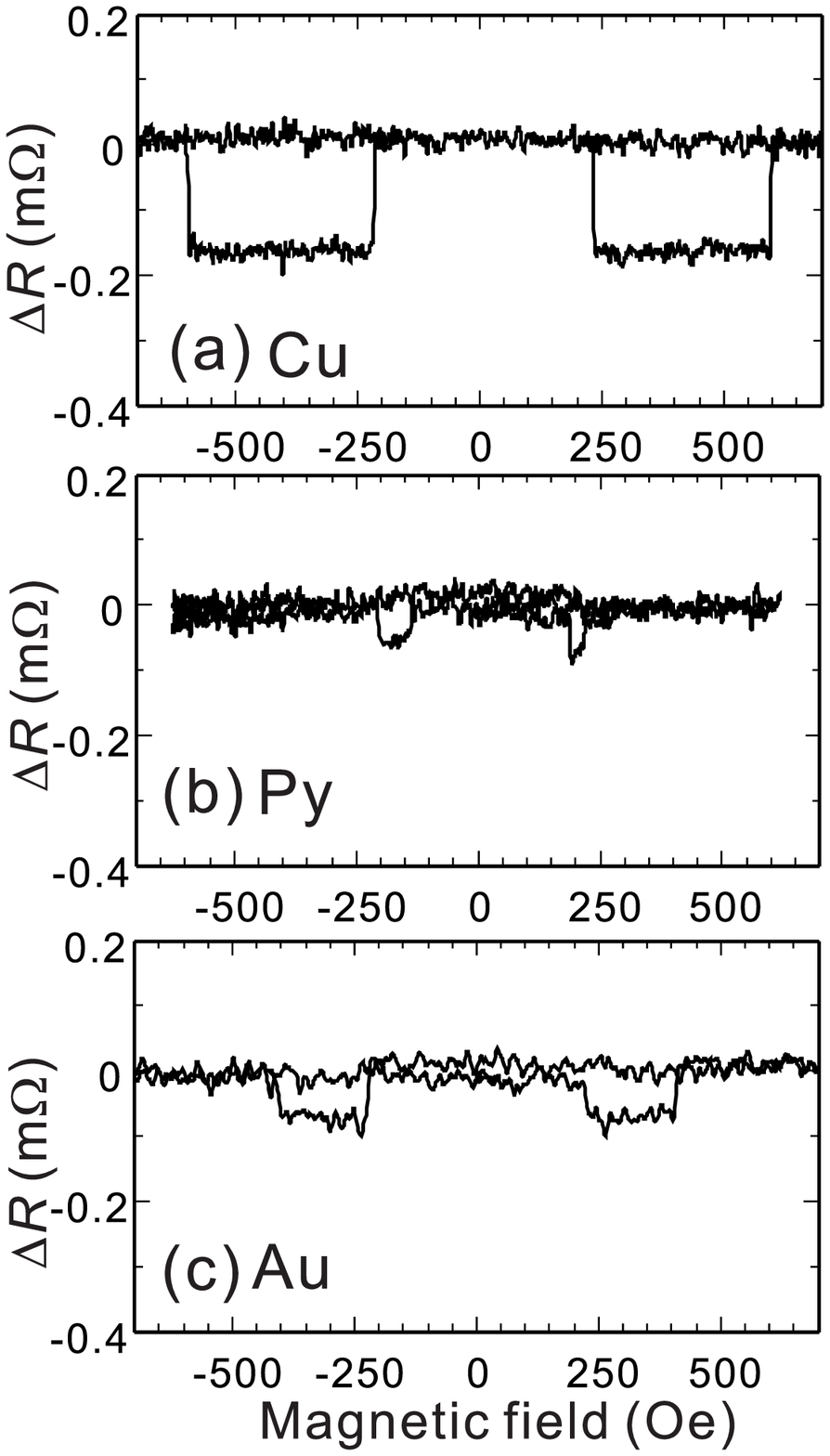}
\end{center}
\vspace*{1cm}
\begin{center}
Fig.\ 5 Kimura et al.
\end{center}

\newpage
\vspace*{4cm}
\begin{center}
\includegraphics[scale=0.8]{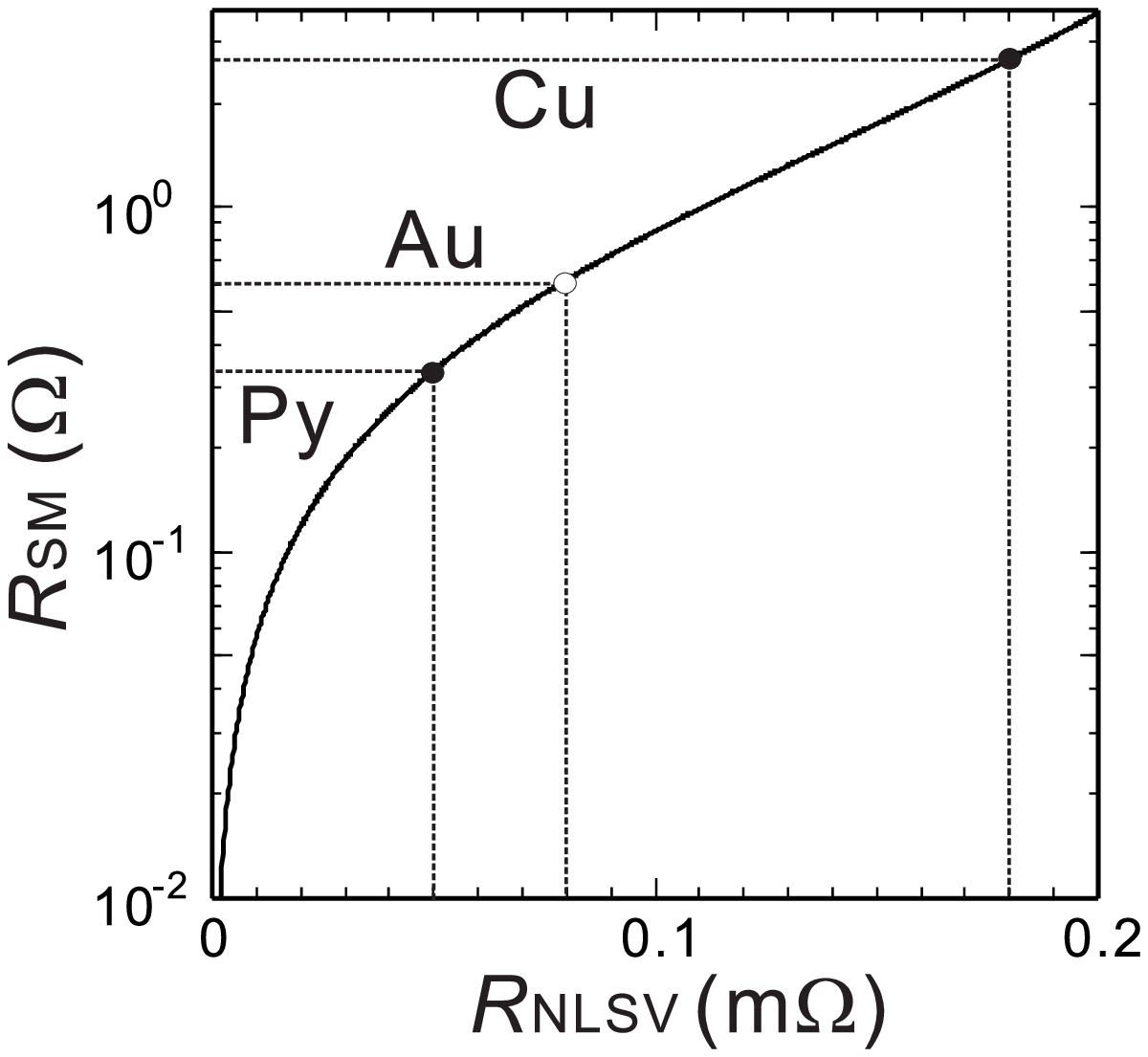}
\end{center}
\vspace*{1cm}
\begin{center}
Fig.\ 6 Kimura et al.
\end{center}

\end{document}